\documentclass[a4,11pt,reqno]{article}
\usepackage[english]{babel}
\usepackage[latin1]{inputenc}
\usepackage{graphics}
\usepackage{graphicx}
\usepackage{ulem}
\usepackage{hhline}
\usepackage{dsfont}
\usepackage{mathrsfs}
\usepackage{fancyhdr}
\usepackage{amsmath,amssymb}
\usepackage{rotating}
\usepackage[latin1]{inputenc}
\usepackage[T1]{fontenc}
\usepackage{fancybox}
\usepackage{color}
\usepackage{colortbl}
\usepackage{setspace}
\usepackage{enumerate}
\usepackage[nice]{nicefrac}
\usepackage{amsthm}
\usepackage{wasysym}
\usepackage{epsfig}
\usepackage{multicol}
\usepackage{pifont}
\usepackage[french]{minitoc}
\usepackage{latexsym,amsfonts}
\usepackage{varioref}
\usepackage{textcomp}
\usepackage{lmodern}
\usepackage{mathpazo}
\usepackage{euscript}
\usepackage[pdftex]{hyperref}

\numberwithin{equation}{section} \makeatletter\@addtoreset{equation}{section}

\DeclareMathSymbol{\subsetneqq}{\mathbin}{AMSb}{36}

\begin{document}

\author{Zouha\"ir MOUAYN}
\title{Coherent state transforms attached to generalized Bargmann spaces
on the complex plane}

\date{{\small\it Department of Mathematics, Faculty of Sciences and Technics
(M'Ghila),}\\
{\small \it Sultan Moulay Slimane University, BP 523, B\'{e}ni Mellal, Morocco }%
\\
{\small (e-mail: mouayn@gmail.com)}}
\maketitle

\begin{abstract}
We construct a family of coherent states transforms attached to generalized
Bargmann spaces [\textit{C.R. Acad.Sci.Paris, t.325,1997}]  in the complex plane. This constitutes another way of obtaining
the kernel of an isometric operator linking the space of square integrable
functions on the real line with the \textit{true-poly}-Fock spaces [\textit{Oper.Theory. Adv.Appl.,v.117,2000}].
\end{abstract}

\section{Introduction}

The Bargmann transform, was originally introduced in 1961 by V. Bargmann
\cite{Bargmann61} and was closely connected to the Heisenberg group. It has
found many applications in quantum optics. Another interest on this transform
lies in that it is a windowed Fourier transform \cite{Hall01}  and as
such it plays an important role in signal processing and harmonic analysis
on phase space \cite{Folland89}.

This transform can be defined as
\begin{equation}
\mathcal{B}\left[ f\right] \left( z\right) :=\pi ^{-\frac{1}{4}}\int\limits_{%
\Bbb{R}}f\left( \xi \right) e^{-\frac{1}{2}\xi ^{2}+\sqrt{2}\xi z-\frac{1}{2}%
z^{2}}d\xi ,z\in \Bbb{C}.  \label{Eq1.1}
\end{equation}
It maps isometrically the space $L^{2}(\Bbb{R},d\xi )$ of square integrable
functions $f$ on the real line onto the Fock space $\frak{F}\left( \Bbb{C}
\right) $ of entire complex-valued functions which are $e^{-\left| z\right|
^{2}}d\lambda -$square integrable, $d\lambda $ denotes the ordinary planar
Lebesgue measure.

Note also that\ the Fock space $\frak{F}\left( \Bbb{C}\right) $
coincides with the null space
\begin{equation}
\mathcal{A}_{0}\left( \Bbb{C}\right) :=\left\{ \varphi \in L^{2}(\Bbb{C}%
,e^{-\left| z\right| ^{2}}d\lambda ), \widetilde{\Delta }\varphi
=0\right\}   \label{Eq1.2}
\end{equation}
\textsf{\ }of the second order differential operator \cite{AsInMo97}:
\begin{equation}
\widetilde{\Delta }:=-\frac{\partial ^{2}}{\partial z\partial \overline{z}}+%
\overline{z}\frac{\partial }{\partial \overline{z}}.  \label{Eq1.3}
\end{equation}
The latter constitutes (in suitable units and up to additive\textsf{\ }%
constant) a\textsf{\ }realization in\textsf{\ }$L^{2}(\Bbb{C},e^{-\left|
z\right| ^{2}}d\lambda )$\textsf{\ }of the Schr\"{o}dinger operator
describing the motion of a charged particle evolving in the complex plane $%
\Bbb{C}$ under influence of a normal uniform magnetic field\textsf{.} Its
spectrum consists of eigenvalues of infinite multiplicity (\textit{Landau
levels}) of the form\textsf{\ :}
\[
\epsilon _{m}=m,\mathsf{\ }m=0,1,2,...
\]
\textsf{\ }The corresponding eigenspaces
\begin{equation}
\mathcal{A}_{m}\left( \Bbb{C}\right) :=\left\{ \varphi \in L^{2}(\Bbb{C}%
,e^{-\left| z\right| ^{2}}d\lambda ), \widetilde{\Delta }\varphi
=\epsilon _{m}\varphi \right\}   \label{Eq1.4}
\end{equation}
\textsf{\ } are pairwise orthogonal in the Hilbert space $L^{2}(\Bbb{C}%
,e^{-\left| z\right| ^{2}}d\lambda )$ which decomposes as
\[
L^{2}(\Bbb{C},e^{-\left| z\right| ^{2}}d\lambda )=\bigoplus\limits_{m\geq 0}%
\mathcal{A}_{m}\left( \Bbb{C}\right) .
\]

In this Note, the main objective is to construct for each Hilbert space $%
\mathcal{A}_{m}\left( \Bbb{C}\right) $, $m=0,1,2,...$ a unitary
transformation, $\mathcal{B}_{m}:L^{2}\left( \Bbb{R}\right) \rightarrow $ $%
\mathcal{A}_{m}\left( \Bbb{C}\right) $ in a such a way that for the first
Hilbert space $\mathcal{A}_{0}\left( \Bbb{C}\right) $, which is the Fock
space, the constructed transform $\mathcal{B}_{0}$ coincides with the
classical Bargmann transform $\mathcal{B}.$ This will be achieved by adopting
a coherent states analysis. Precisely, the constructed transforms are of the
form
\[
\mathcal{B}_{m}\left[ f\right] \left( z\right) =\left( -1\right) ^{m}\left(
2^{m}m!\sqrt{\pi }\right) ^{-\frac{1}{2}}\int\limits_{\Bbb{R}}f\left( \xi
\right) e^{-\frac{1}{2}\xi ^{2}+\sqrt{2}\xi z-\frac{1}{2}z^{2}}H_{m}\left(
\xi -\frac{z+\overline{z}}{2}\right) d\xi ,
\]
where $H_{m}(\xi )$\textit{\ }$=\left( -1\right) ^{m}e^{\xi ^{2}}\left(
\frac{d}{d\xi }\right) ^{m}e^{-\xi ^{2}}$is the\textit{\ }$m$th Hermite
polynomial.

We should note that the expression of the transforms $\mathcal{B}_{m}$
coincides with the expression of a family of isometric operators linking the
space $L^{2}\left( \Bbb{R}\right) $ with the \textit{true-poly}-\textit{Fock}
spaces introduced by N. L. Vasilevski \cite{Vasilevski00} . Thereby, the
present work constitutes another way to arrive at the result of theorem 2.5
in \cite{Vasilevski00}, by using a coherent states method exploiting tools of
the $L^{2}-$spectral theory of the Schr\"{o}dinger operator given in (1.3) .

In the next section, We review briefly the coherent states formalism we will
be using. Section 3 deals with some needed facts on the $L^{2}-$spectral
theory of the Schr\"{o}dinger operator $\widetilde{\Delta }$. In section 4
we define a family of coherent state transforms attached to the generalized
Bargmann spaces $\mathcal{A}_{m}\left( \Bbb{C}\right) $.

\section{Coherent states formalism}

Here, we follow the generalization of the canonical coherent states
according to the procedure in \cite{ThiruSaad04}.

Let $(X,\mu )$\ be a measure space and let $\frak{H}^{2}\subset L^{2}(X,\mu
) $\ be a closed subspace of infinite dimension. Let $\left\{ \Phi
_{n}\right\} _{n=0}^{\infty }$ be an orthogonal basis of $\frak{H}^{2}$
satisfying, for arbitrary $x\in X,$
\[
\omega \left( x\right) :=\sum_{n=0}^{\infty }\rho _{n}^{-1}\left| \Phi
_{n}\left( x\right) \right| ^{2}<+\infty ,
\]
where $\rho _{n}:=\left\| \Phi _{n}\right\| _{L^{2}(X)}^{2}$. Define
\[
\frak{K}(x,y):=\sum_{n=0}^{\infty }\rho _{n}^{-1}\Phi _{n}\left( x\right)
\overline{\Phi _{n}(y)},~x,y\in X.
\]
Then, $\frak{K}(x,y)$\ is a reproducing kernel, $\frak{H}^{2}$ is the
corresponding reproducing kernel Hilbert space and $\omega \left( x\right) :=%
\frak{K}(x,x)$, $x\in X$.

\textbf{Definition 2.1}. \textit{Let }$\mathcal{H}$\textit{\ be a Hilbert
space with }$\dim \mathcal{H}=\infty $\textit{\ and }$\left\{ \phi
_{n}\right\} _{n=0}^{\infty }$\textit{\ be an orthonormal basis of }$%
\mathcal{H}.$\textit{\ The coherent states labeled by points }$x\in X$%
\textit{\ are defined as the ket-vectors }$\vartheta _{x}\equiv \mid x>\in
\mathcal{H}:$%
\begin{equation}
\vartheta _{x}\equiv \mid x>:=\left( \omega \left( x\right) \right) ^{-\frac{%
1}{2}}\sum_{n=0}^{\infty }\frac{\Phi _{n}\left( x\right) }{\sqrt{\rho _{n}}}%
\phi _{n}.\quad \quad  \label{Eq2.1}
\end{equation}
By definition, it is straightforward to show that $<\vartheta _{x},\vartheta
_{x}>_{\mathcal{H}}=1.$

\textbf{Definition 2.2.} \textit{\ The coherent state transform associated
to the set of coherent states }$\left( \vartheta _{x}\right) _{x\in X}$%
\textit{\ is the isometric map}
\begin{equation}
W:\mathcal{H}\rightarrow \frak{H}^{2}\subset L^{2}(X,\mu )  \label{Eq2.2}
\end{equation}
\textit{defined for every }$x\in X$ \textit{by}
\[
W\left[ \phi \right] \left( x\right) :=\left( \omega \left( x\right) \right)
^{\frac{1}{2}}<\phi ,\vartheta _{x}>_{\mathcal{H}}.
\]
Thus, for $\phi ,\psi \in \mathcal{H}$, we have
\[
<\phi ,\psi >_{\mathcal{H}}=<W\left[ \phi \right] ,W\left[ \psi \right]
>_{L^{2}\left( X\right) }=\int\limits_{X}d\mu \left( x\right) \omega \left(
x\right) <\phi ,\vartheta _{x}><\vartheta _{x},\psi >.
\]
Thereby, we have a resolution of the identity of $\mathcal{H}$ which can be
expressed in Dirac's bra-ket notation as:
\[
\mathbf{1}_{\mathcal{H}}=\int\limits_{X}d\mu \left( x\right) \omega \left(
x\right) \mid x><x\mid ,\quad \quad
\]
and where $\omega \left( x\right) $\ appears as a weight function.

\textbf{Remark 2.1}. Note that formula  \eqref{Eq2.1} can be considered as a
generalization of the series expansion of the canonical coherent states
\[
\vartheta _{\zeta }\equiv \mid \zeta >:=e^{-\frac{1}{2}\left| \zeta \right|
^{2}}\sum_{k=0}^{+\infty }\frac{\zeta ^{k}}{\sqrt{k!}}\phi _{k},\zeta \in
\Bbb{C}
\]
with $\left\{ \phi _{k}\right\} _{k=0}^{+\infty }$\ being an orthonormal
basis of eigenstates of the quantum harmonic oscillator. Here, the space $%
\frak{H}^{2}$ is the Fock space $\frak{F}\left( \Bbb{C}\right) $ and $\omega
(\zeta )=\pi ^{-1}e^{\left| \zeta \right| ^{2}},\zeta \in \Bbb{C}$.

\section{The generalized Fock spaces $\mathcal{A}_{m}\left( \Bbb{C}\right) $}

As the Fock space $\frak{F}\left( \Bbb{C}\right) $ has $K_{0}\left(
z,w\right) :=\pi ^{-1}e^{z\overline{w}}$ as reproducing kernel, we have
shown \cite{AsInMo97}  that the Hilbert spaces $\mathcal{A}_{m}\left( \Bbb{%
C}\right) $ also have explicit reproducing kernel of the form
\begin{equation}
K_{m}(z,w):=\pi ^{-1}e^{\left\langle z,w\right\rangle }L_{m}^{\left(
0\right) }\left( \left| z-w\right| ^{2}\right) ,z,w\in \Bbb{C},  \label{Eq3.1}
\end{equation}
where $L_{m}^{\left( \alpha \right) }\left( t\right) $ is the Laguerre
polynomial defined by the Rodriguez formula as
\[
L_{m}^{\left( \alpha \right) }\left( t\right) =\frac{1}{m!}t^{-\alpha
}e^{t}\left( \frac{d}{dt}\right) ^{m}\left( t^{\alpha +m}e^{-t}\right) ,t\in
\Bbb{R}
\]
In particular, if we set $\omega _{m}\left( z\right) :=K_{m}(z,z),$ then $%
\omega _{m}\left( z\right) =\pi ^{-1}e^{\left| z\right| ^{2}},$ $z\in \Bbb{C}%
.$

The spaces $\mathcal{A}_{m}\left( \Bbb{C}\right) $ have been also used to
study the spectral properties of the Cauchy transform on $L^{2}(\Bbb{C}%
,e^{-\left| z\right| ^{2}}d\lambda );$ see \cite{InIn06}  where the
authors exhibited for each fixed $m=0,1,2,...$ an orthogonal basis denoted $%
\left\{ h_{m,p}\right\} _{p=0}^{+\infty }$ and defined by
\begin{equation}
h_{m,p}\left( z\right) :=\gamma _{m,p\text{ \ }1}\digamma _{1}\left( -\min
\left( m,p\right) ,\left| m-p\right| +1,\left| z\right| ^{2}\right) \left|
z\right| ^{\left| m-p\right| }e^{-i(m-p)\arg z}  \label{Eq3.2}
\end{equation}
where
\[
\gamma _{m,p}:=\frac{\left( -1\right) ^{\min \left( m,p\right) }\left( \max
\left( m,p\right) \right) !}{\left( \left| m-p\right| \right) !},
\]
and $_{1}\digamma _{1}$ is the confluent hypergeometric function given by  \cite{GradRyz80}:
\[
_{1}\digamma _{1}\left( a,b;u\right) =\frac{\Gamma \left( b\right) }{\Gamma
\left( a\right) }\sum\limits_{j=0}^{+\infty }\frac{\Gamma \left( a+j\right)
}{\Gamma \left( b+j\right) }\frac{u^{j}}{j!},\left| u\right| <+\infty ,b\neq
0,-1,-2,...\text{ .}
\]
Here $\Gamma \left( a\right) $ is the Euler's Gamma function such that $%
\Gamma \left( j+1\right) =j!$ if $j=0,1,2,....$

Note that for $a=-n$ with $n$ being a positive integer, the hypergeometric
function $_{1}\digamma _{1}$ becomes a polynomial and can be expressible in
term of Laguerre polynomial according to \cite{GradRyz80}:
\[
_{1}\digamma _{1}\left( -n,\alpha +1;u\right) =\frac{n!\Gamma \left( \alpha
+1\right) }{\Gamma \left( n+\alpha +1\right) }L_{n}^{\left( \alpha \right)
}\left( u\right) .
\]
For our purpose we shall consider the orthogonal basis of $\mathcal{A}%
_{m}\left( \Bbb{C}\right) $ in the following form
\begin{equation}
h_{m,p}\left( z\right) =\left( -1\right) ^{\min \left( m,p\right) }\left(
\min (m,p\right) )!\left| z\right| ^{\left| m-p\right| }e^{-i(m-p)\arg
z}L_{\min \left( m,p\right) }^{(\left| m-p\right| )}\left( \left| z\right|
^{2}\right) ,z\in \Bbb{C},  \label{Eq3.3}
\end{equation}
with the square norm in $L^{2}(\Bbb{C},e^{-\left| z\right| ^{2}}d\lambda )$
given by
\[
\rho _{m,p}:=\left\| h_{m,p}\right\| ^{2}=\pi m!p!.
\]
\textbf{Remark 3.1. }In \cite[p. 404]{InIn06} the elements
of the orthogonal basis given in \eqref{Eq3.2}  have been also
expressed as
\begin{equation}
h_{m,p}\left( z\right) =\sum\limits_{j=0}^{\min \left( m,p\right) }\left(
-1\right) ^{j}\frac{m!p!}{j!\left( m-j\right) !\left( p-j\right) !}z^{m-j}%
\overline{z}^{p-j}.  \label{Eq3.4}
\end{equation}
We should note these complex polynomials in $\left( 3.4\right) $ were
considered also by It\^{o} \cite{Ito53}  in the context of complex
Markov process.

\section{Coherent states attached to $\mathcal{A}_{m}\left( \Bbb{C}\right) $}

In this section, we shall attach to each space $\mathcal{A}_{m}\left( \Bbb{C}%
\right) $ a set coherent states via series expansion according to the
procedure presented in section 2. We will also give expressions of these
coherent states in a closed form by using direct calculations.

\textbf{Definition 4.1. }\textit{For }$m=0,1,2,...$., \textit{the coherent
states associated with the space }$\mathcal{A}_{m}\left( \Bbb{C}\right) $%
\textit{\ and labelled by points }$z\in \Bbb{C}$\textit{\ are defined
formally according to formula \eqref{Eq2.1} as}
\[
\vartheta _{z,m}\equiv \mid z,m>:=\left( \omega _{m}(z)\right) ^{-\frac{1}{2}%
}\sum_{p=0}^{+\infty }\frac{h_{m,p}\left( z\right) }{\sqrt{\rho _{m,p}}}\psi
_{p}
\]
\textit{where }$\psi _{p}$\textit{\ \ are elements of a total orthonormal
system of }$L^{2}(\Bbb{R},d\xi )$\textit{\ given }
\[
\psi _{p}(\xi ):=\left( \sqrt{\pi }2^{p}p!\right) ^{-\frac{1}{2}}e^{-\frac{1%
}{2}\xi ^{2}}H_{p}(\xi ),~p=0,1,2,...,\quad \xi \in \Bbb{R}\text{,}
\]
\textit{and} $H_{p}(\xi )$ \textit{is the }$p$\textit{th Hermite polynomial.
}

\textbf{Proposition 4.1. }\textit{The}\textbf{\ }\textit{wave functions of
these coherent states are expressed as}
\[
\vartheta _{z,m}\left( \xi \right) =\left( -1\right) ^{m}\left( 2^{m}m!\sqrt{%
\pi }\right) ^{-\frac{1}{2}}e^{-\frac{1}{2}\overline{z}^{2}+\sqrt{2}\xi
\overline{z}-\frac{1}{2}\left| z\right| ^{2}-\frac{1}{2}\xi ^{2}}H_{m}\left(
\xi -\frac{z+\overline{z}}{2}\right) ,\xi \in \Bbb{R}\text{.}
\]

\textbf{Proof.} According  to Definition 4.1, we start by writing
\[
\vartheta _{z,m}\left( \xi \right) =\left( \frac{1}{\pi }e^{\left| z\right|
^{2}}\right) ^{-\frac{1}{2}}\sum_{p=0}^{+\infty }\frac{h_{m,p}\left(
z\right) }{\sqrt{\pi m!p!\ }}\psi _{p}\left( \xi \right) .
\]
Recalling the expression of $h_{m,p}\left( z\right) $ in $\left( 3.3\right) $%
, then these wave functions can be rewritten as
\[
\vartheta _{z,m}\left( \xi \right) =\frac{e^{-\frac{1}{2}\left| z\right|
^{2}}}{\sqrt{m!}}\sum_{p=0}^{+\infty }\frac{\left( -1\right) ^{\min \left(
m,p\right) }}{\sqrt{p!}}\left( \min \left( m,p\right) \right) !\left|
z\right| ^{\left| m-p\right| }e^{-i(m-p)\arg z}L_{\min \left( m,p\right)
}^{(\left| m-p\right| )}\left( \left| z\right| ^{2}\right) \psi _{p}\left(
\xi \right) .
\]
The integer $m$ being fixed, we denote by $\frak{S}_{m}\left( z,\xi \right) $
the following series:
\[
\frak{S}_{m}\left( z,\xi \right) :=\sum_{p=0}^{+\infty }\frac{\left(
-1\right) ^{\min \left( m,p\right) }}{\sqrt{p!}}\left( \min (m,p)\right)
!\left| z\right| ^{\left| m-p\right| }e^{-i(m-p)\arg z}L_{\min \left(
m,p\right) }^{(\left| m-p\right| )}\left( \left| z\right| ^{2}\right) \psi
_{p}\left( \xi \right)
\]
and we split it into two part as
\begin{eqnarray*}
\frak{S}_{m}\left( z,\xi \right) &=&\sum_{p=0}^{m-1}\frac{1}{\sqrt{p!}}%
\left( -1\right) ^{p}p!\left| z\right| ^{m-p}e^{-i(m-p)\arg
z}L_{p}^{(m-p)}\left( \left| z\right| ^{2}\right) \psi _{p}\left( \xi \right)
\\
&&+\sum_{p=m}^{+\infty }\frac{1}{\sqrt{p!}}\left( -1\right) ^{m}m!\left|
z\right| ^{p-m}e^{-i(m-p)\arg z}L_{m}^{(p-m)}\left( \left| z\right|
^{2}\right) \psi _{p}\left( \xi \right)
\end{eqnarray*}
This can also be written as
\[
\frak{S}\left( m,z,\xi \right) =\mathcal{S}_{(<\infty )}\left( m,z,\xi
\right) +\mathcal{S}_{(\infty )}\left( m,z,\xi \right)
\]
with
\begin{eqnarray*}
\mathcal{S}_{(<\infty )}\left( m,z,\xi \right) &=&\sum_{p=0}^{m-1}\frac{1}{%
\sqrt{p!}}\left( -1\right) ^{p}p!\left| z\right| ^{m-p}e^{-i(m-p)\arg
z}L_{p}^{(m-p)}\left( \left| z\right| ^{2}\right) \psi _{p}\left( \xi \right)
\\
&&-\sum_{p=0}^{m-1}\frac{1}{\sqrt{p!}}\left( -1\right) ^{m}m!\left| z\right|
^{p-m}e^{-i(m-p)\arg z}L_{m}^{(p-m)}\left( \left| z\right| ^{2}\right) \psi
_{p}\left( \xi \right)
\end{eqnarray*}
and
\[
\mathcal{S}_{(\infty )}\left( m,z,\xi \right) =\sum_{p=0}^{+\infty }\frac{1}{%
\sqrt{p!}}\left( -1\right) ^{m}m!\left| z\right| ^{p-m}e^{-i(m-p)\arg
z}L_{m}^{(p-m)}\left( \left| z\right| ^{2}\right) \psi _{p}\left( \xi
\right) .
\]
The finite sum $\mathcal{S}_{(<\infty )}\left( m,z,\xi \right) $ reads
\[
\mathcal{S}_{(<\infty )}\left( m,z,\xi \right) =\sum_{p=0}^{m-1}\left(
\left( -1\right) ^{p}\sqrt{p!}\overline{z}^{m-p}L_{p}^{(m-p)}\left( \left|
z\right| ^{2}\right) -\left( -1\right) ^{m}\frac{m!}{\sqrt{p!}}%
z^{p-m}L_{m}^{(p-m)}\left( \left| z\right| ^{2}\right) \right) \psi
_{p}\left( \xi \right)
\]
Making use of the identity  \cite[p. 98]{Szego75}:
\[
L_{m}^{\left( -k\right) }\left( t\right) =\left( -t\right) ^{k}\frac{\left(
m-k\right) !}{m!}L_{m-k}^{\left( k\right) }\left( t\right) ,~1\leq k\leq m
\]
for $k=p-m,$ we write the Laguerre polynomial with upper indice $p-m<0$ as
\[
L_{m}^{(p-m)}\left( \left| z\right| ^{2}\right) =\left( -\left| z\right|
^{2}\right) ^{m-p}\frac{p!}{m!}L_{p}^{(m-p)}\left( \left| z\right|
^{2}\right) ,
\]
and we obtain after calculation that $\mathcal{S}_{(<\infty )}\left( m,z,\xi
\right) =0.$

Now, for the infinite sum $\mathcal{S}_{(\infty )}\left( m,z,\xi \right) ,$
we make use of the explicit expression of the Gaussian-Hermite functions
\[
\psi _{p}\left( \xi \right) =\left( \sqrt{\pi }2^{p}p!\right) ^{-\frac{1}{2}%
}e^{-\frac{1}{2}\xi ^{2}}H_{p}\left( \xi \right) ,p=0,1,2,...
\]
and we obtain that
\begin{eqnarray*}
\mathcal{S}_{(\infty )}\left( m,z,\xi \right) &=&\sum_{p=0}^{+\infty }\frac{1%
}{\sqrt{p!}}\left( -1\right) ^{m}m!\left| z\right| ^{p-m}e^{-i(m-p)\arg
z}L_{m}^{(p-m)}\left( \left| z\right| ^{2}\right) \frac{e^{-\frac{1}{2}\xi
^{2}}H_{p}\left( \xi \right) }{\left( \sqrt{\pi }2^{p}p!\right) ^{\frac{1}{2}%
}} \\
&=&\left( \sqrt{\pi }\right) ^{-\frac{1}{2}}\left( -1\right) ^{m}m!e^{-\frac{%
1}{2}\xi ^{2}}\frak{T}_{\left( \infty \right) }\left( m,z,\xi \right)
\end{eqnarray*}
where
\[
\frak{T}_{\left( \infty \right) }\left( m,z,\xi \right)
:=\sum_{p=0}^{+\infty }\frac{\left( 2^{p}\right) ^{-\frac{1}{2}}}{p!}%
z^{p-m}L_{m}^{(p-m)}\left( \left| z\right| ^{2}\right) H_{p}\left( \xi
\right)
\]
Next, we make use of following addition formula involving Laguerre and
Hermite polynomials \cite{Submited}:
\begin{eqnarray*}
&&\sum_{j=-n}^{+\infty }\frac{2^{-j}\beta ^{\frac{j}{2}}}{\left( j+n\right) !%
}\left( a+ib\right) ^{j}L_{n}^{(j)}\left( \frac{\beta }{2}\left(
a^{2}+b^{2}\right) \right) H_{j+n}\left( \xi \right) \\
&=&\frac{1}{m!}\exp \left( -\frac{\beta }{4}\left( a-ib\right) ^{2}+\sqrt{%
\beta }\xi \left( a-ib\right) \right) H_{n}\left( \xi -\sqrt{\beta }a\right)
\end{eqnarray*}
for $n=m,$ $j=p-n,$ $\beta =2$ and $z=a+ib\in \Bbb{C}.$ This gives that
\[
\frak{T}_{\left( \infty \right) }\left( m,z,\xi \right) =\frac{2^{-\frac{m}{2%
}}}{m!}e^{-\frac{1}{2}\overline{z}^{2}+\sqrt{2}\xi \overline{z}}H_{n}\left(
\xi -\frac{z+\overline{z}}{2}\right)
\]
Summarizing up the above calculations, we can write successively
\begin{eqnarray*}
\vartheta _{z,m}\left( \xi \right) &=&\frac{e^{-\frac{1}{2}\left| z\right|
^{2}}}{\sqrt{m!}}\left( \sqrt{\pi }\right) ^{-\frac{1}{2}}\left( -1\right)
^{m}m!e^{-\frac{1}{2}\xi ^{2}}\frak{T}_{\left( \infty \right) }\left(
m,z,\xi \right) \\
&=&\frac{e^{-\frac{1}{2}\left| z\right| ^{2}}}{\sqrt{m!}}\left( \sqrt{\pi }%
\right) ^{-\frac{1}{2}}\left( -1\right) ^{m}m!e^{-\frac{1}{2}\xi ^{2}}\left(
\frac{2^{-\frac{m}{2}}}{m!}e^{-\frac{1}{2}\overline{z}^{2}+\sqrt{2}\xi
\overline{z}}H_{m}\left( \xi -\frac{z+\overline{z}}{2}\right) \right) \\
&=&\left( -1\right) ^{m}\left( 2^{m}m!\sqrt{\pi }\right) ^{-\frac{1}{2}}e^{-%
\frac{1}{2}\overline{z}^{2}+\sqrt{2}\xi \overline{z}-\frac{1}{2}\left|
z\right| ^{2}-\frac{1}{2}\xi ^{2}}H_{m}\left( \xi -\frac{z+\overline{z}}{2}%
\right) .
\end{eqnarray*}
The proof of Proposition 4.1 is finished. $\blacksquare $

Finally, according to Definition 2.2, the coherent state transform
associated with the coherent states $\vartheta _{z,m}$ is the unitary map:
\[
\mathcal{B}_{m}:L^{2}(\Bbb{R},d\xi )\rightarrow \mathcal{A}_{m}\left( \Bbb{C}%
\right)
\]
defined by
\[
\mathcal{B}_{m}\left[ f\right] \left( z\right) :=\left( \omega _{m}\left(
z\right) \right) ^{\frac{1}{2}}\left\langle f,\vartheta _{z,m}\right\rangle
_{L^{2}(\Bbb{R})},f\in L^{2}(\Bbb{R},d\xi ),z\in \Bbb{C}
\]
\ Explicitly,
\[
\mathcal{B}_{m}\left[ f\right] \left( z\right) =\left( -1\right) ^{m}\left(
2^{m}m!\sqrt{\pi }\right) ^{-\frac{1}{2}}\int\limits_{\Bbb{R}}f\left( \xi
\right) e^{-\frac{1}{2}\xi ^{2}+\sqrt{2}\xi z-\frac{1}{2}z^{2}}H_{m}\left(
\xi -\frac{z+\overline{z}}{2}\right) d\xi
\]
which can be called the extended Bargmann transform of index $m=0,1,2,...$.


\end{document}